\documentclass[12pt]{article}

\input epsf
\def\beq{\begin{eqnarray}}
\def\eeq{\end{eqnarray}}

\def\lsim{\mathrel{\rlap{\lower3pt\hbox{\hskip0pt$\sim$}}
    \raise1pt\hbox{$<$}}}         
\def\gsim{\mathrel{\rlap{\lower4pt\hbox{\hskip1pt$\sim$}}
    \raise1pt\hbox{$>$}}}         

\begin{document}


\vskip 1cm
\begin{center}
{\Large \bf Cosmic Attractors and Gauge Hierarchy}
\vskip 1cm
{ Gia Dvali$^1$ and Alexander Vilenkin$^2$}

\vskip 1cm
{\it  $^1$ Center for Cosmology and Particle Physics, Department of Physics, New York University, New York, NY 10003\\
$^2$ Institute of Cosmology, Department of Physics and Astronomy,\\
Tufts University, Medford, MA 02155, USA}
\end{center}

\vspace{0.9cm}
\begin{center}
{\bf Abstract}
\end{center}

We suggest a new cosmological scenario which naturally guarantees the
smallness of scalar masses and VEVs, without invoking supersymmetry or
any other (non-gravitationaly coupled) new physics at low energies. 
 In our framework, the scalar
masses undergo discrete jumps due to nucleation of closed branes
during (eternal) inflation. The crucial point is that the step size of
variation decreases in the direction of decreasing scalar mass.
This scenario yields exponentially large domains with a distribution of
scalar masses, which is sharply peaked around a hierarchically small
value of the mass. This value is the "attractor point" of the
cosmological evolution.

\vspace{0.1in}


\section{General idea}

The radiative instability of scalar masses is the key point of the
gauge hierarchy problem. In the effective 4D field theory, the scalar
masses are quadratically sensitive to the ultraviolet cutoff.  The
only known exceptions to this rule are Goldstone bosons.  This fact is
hard to reconcile with the observed smallness of the weak scale,
relative to the Planck mass $M_p \, \sim \, 10^{19} $GeV.  So far
supersymmetry is the only known symmetry that renders masses of
elementary scalars radiatively stable. The scalar masses are
controlled by supersymmetry breaking scale.  Given the fact that we do
not understand the origin of this scale, supersymmetry {\it per se}
does not really explain the origin of the weak scale but rather makes
the gauge hierarchy technically natural.

In view of the above, it is crucial to explore other possible
mechanisms of scalar mass stabilization.  In the present paper we
suggest an alternative mechanism that can guarantee zero or very small
scalar masses (and VEVs) without invoking supersymmetry or any other
non-gravitationaly-coupled new physics at low energies.

In our scenario, a small scalar mass is selected with probability one
during the cosmological evolution.  This selection works as follows.
We construct a simple framework in which scalar masses (and VEVs)
undergo discrete variations due to nucleation of closed domain wall
bubbles (branes) during inflation.  Values of the scalar mass on
different sides of the wall differ by a finite step.  The bubbles
expand exponentially fast and create domains of a new vacuum with a
new value of the scalar mass.  New bubbles are created within the old,
and the scalar mass changes further.  Since inflation is known to be
eternal \cite{AV83,Linde86}, the process of wall nucleation continues
forever, populating the Universe with exponentially large domains
having different values of the scalar mass.  However, not all the
values of the scalar mass (VEV) are equally probable.  In our model,
in the absence of gravitational back-reaction, the probability is
sharply peaked around zero, because the step $\Delta\phi$  decreases
towards small values of the VEV $\phi$ faster than
$\phi$ itself.  That is,
\begin{equation}
\label{decrease}
\Delta \phi /\phi \, \propto \phi^n,
\end{equation}
where $n>0$ is some power.  As a result, the density of states diverges 
for small VEV (mass) of $\phi$. 



Thus, in the first approximation, the probability distribution for
$\phi$ has an infinitely sharp peak at $\phi \, = \, 0$.  We will
show, however, that infrared effects, such as the Gibbons-Hawking
temperature and quantum fluctuation of $\phi$ during inflation, can
shift the most probable value of the scalar mass (and VEV) away from
zero to a small value and round off the maximum of the peak.

\section{Cosmic attractors}

To introduce our mechanism, we use a simple toy model. The main
ingredients are: (1) a scalar field $\phi$; (2) domain walls
(branes) charged under an antisymmetric three-form field
$A_{\alpha\beta\gamma}$, with the field-strength
$F_{\alpha\beta\gamma\delta}=F\epsilon_{\alpha\beta\gamma\delta}$.   
These objects are engaged in the following
interrelation. The branes are sources for the three-form field.  The
value of the brane charge is determined by the VEV of $\phi$.  The VEV of
$\phi$ is in turn determined by the three-form field strength $F$.

These couplings result in the following dynamics.  Nucleation of a
closed brane changes the value of $F$. The step of
change (the brane charge) is determined by $\phi$.  We construct the
model so that an increase in $F$ decreases $\phi$, which in turn
decreases the charge of new branes that can be nucleated.  Decrease of
the brane charge diminishes the minimal step of change in $F$.  As a
result, the subsequent decrease of $\phi$ requires more steps, and
their number diverges towards small values of $\phi$.

Let us discuss this dynamics in more detail.  The action of a free
three-form field in 4D can be written as
\begin{equation}
\label{faction}
\int_{3+1}F^2.
\end{equation}
It is invariant under gauge transformations
\begin{equation}
\label{gauge}
A_{\alpha\beta\gamma} \, \rightarrow \, A_{\alpha\beta\gamma}\, + \,
\partial_{[\alpha}B_{\beta\gamma]},
\end{equation}
where $B$ is a two-form.  Due to this gauge freedom, $F$ contains no
propagating degrees of freedom. The solution to the equations of motion
is an arbitrary constant value of the field strength,
\begin{equation}
\label{solution}
F \, =\, {\rm constant}.
\end{equation}

The situation changes in the presence of 2-branes, or domain walls, which
may act as sources for $A$, due to the following coupling
\begin{equation}
\label{braneA}
q\, \int_{2 + 1} \, A,
\end{equation}
where the integral is taken over the $2+1$-dimensional world-volume and
$q$ is the brane charge.  The role of such branes can be
played by the field-theoretic solitonic domain walls \cite{DV} (see
Appendix A), or by fundamental branes of some sort.  Their precise
origin is unimportant for the present discussion.  The change of $F$
across the wall is given by
\begin{equation}
\label{dF}
\Delta \, F \, = \, q
\end{equation}
Thus, $F$ can undergo discrete variations due to nucleation of closed
branes \cite{teitelboim}.  

This mechanism can be used to induce spatial variations of
the field $\phi$, by coupling it to the $F$-form.  We {\it do not}
require that the Lagrangian contains any small scale (such as the
supersymmetry breaking scale).  We allow the (renormalized) potential
of $\phi$ to be the most general function, including all possible
interactions with the 4-form field $F$,
\begin{equation}
\label{vpf}
V(\phi) \, = \, \left (\,-\, m^2 \, + \, {F^2 \over M_p^2} \, + ...\right )\, 
|\phi|^2\, + \, 
\left ( \, 1\, + \, {F^2 \over M_p^4}\, +.. \right ) |\phi|^4 \, +...
\end{equation} 
The dimensionless coefficients are not
shown explicitly and are assumed to be of order one.  The couplings linear in $F$
are suppressed by parity symmetry. 
The couplings in
(\ref{vpf}) effectively convert the mass and the VEV of $\phi$ into 
functions of the 4-form field strength, e.g.,
\begin{equation}
\label{vevp}
\phi^2 \, \sim \, ( m^2 \, -\,  F^2/ M_p^2)
\end{equation}

We shall assume that the $F$-independent part of the mass $m^2$ takes
its natural value, $m^2 \, \sim \, M_P^2$.  For definiteness, we shall
assume the sign of this contribution to be negative and the sign of
the $F$-dependent contribution to be positive. Then, the $F$-dependent
contribution will lead to a partiall cancellation of the effective
mass.  This mass will take different values in different parts of the
Universe, due to nucleation of branes charged under $F$.  

To ensure that the brane charge $q$ is suppressed at small values
of $\phi$, we require that the system is invariant under a $Z_{2N}$
symmetry, which acts on $\phi$ as
\begin{equation}
\label{zn}
\phi \, \rightarrow \, {\rm e}^{i{\pi \over N}}\, \phi,
\end{equation}
and at the same time changes branes into anti-branes and vice versa
(leaving the 3-form $A$ invariant).\footnote{Alternatively, we could
require that $A \rightarrow -A$, while branes are unaffected.  In fact
the system is invariant under both assignments, independently.}  The
coupling of the 3-form to the branes (\ref{braneA}) should then be
replaced by
\begin{equation}
\label{braneAphi}
 \int_{2 + 1} \, {\phi^N \over M_p^{N-2}} \, A .
\label{phiNA}
\end{equation}
Note that for a non-constant $\phi$ the above coupling is not
gauge-invariant, and extra non-local, terms
have to be added to the action to restore the gauge invariance. A
detailed discussion is given in Appendix A.  These additional terms
are particularly important for understanding of the radiation of
$A$-waves from the branes that occurs if the background value of
$\phi$ changes in time (see Appendix B).  We have shown in
\cite{DV} that a non-local action of this type can arise from a
local field theory in a two-dimensional toy model after integration
over some massless fermions.  It is not clear whether or not this
mechanism can be extended to $4D$, and more generally, whether or not
the required type of action can be obtained in effective field
theory.  We leave this question for future investigation.

In the context of brane nucleation, however, the additional
terms in (\ref{phiNA}) are unimportant.  In the regime of interest to
us here (small $\phi$), each act of brane-nucleation changes $\phi$ by
a very small amount, so $\phi$ can be regarded as nearly constant, and
additional terms are negligible.

The magnitude of the $F$-step between neighboring domains is set by
$\phi$,
\begin{equation}
\label{delF}
\Delta\, F \, \propto \, \phi^N,
\end{equation}
and so is the change of the VEV of $\phi$,
\begin{equation}
\label{delp}
\Delta \, \phi^2 \, \propto \, \phi^N.
\end{equation}
This is the key point of our mechanism.  With every step that
decreases the VEV of $\phi$, we create a region in the Universe where
the brane charges are smaller.  This allows for finer and finer
adjustment of the $\phi$-VEV, accompanied by further decrease of the brane
charges.  Thus, the step size of the field $F$ (and therefore of
$\phi$) decreases and the ``level density'' grows towards smaller
values of $\phi$, and if the process is not for some reason
terminated, the total number of levels diverges.

For instance, imagine that we start in a domain where $F\, \sim \,
1$, and $\phi^2\, = \, \xi \, < 1$ in Planck units.  In these units,
the value of the brane charge in that domain is
\begin{equation}
\label{qcharge}
q_0 \, = \,\xi^{N/2} 
\end{equation}
For the sake of definiteness, let us assume that $\xi \sim 0.1$ or so. 
After the first step  of brane nucleation, the change in $F$ is
\begin{equation}
\label{1step}
\Delta F \, = \, q_0\, \sim  \xi^{N/2}
\end{equation}
and the VEV of $\phi$ is partially cancelled to
\begin{equation}
\label{p1}
\phi^2 \rightarrow (\xi \, - \,\xi^{N/2})
\end{equation}
Then it will take approximately $n \, = \, 1/\xi^{(N/2)-1}$
steps to cancel $\phi^2$ to $\phi^2 \sim \xi^2$. At this point the
brane charge becomes
\begin{equation}
\label{qnew}
q_{new} \, \sim \, \xi^{N },
\end{equation}
and now it will take $n \, \sim \, 1/\xi^{N -2}$ steps to
calcel $\phi$ to
\begin{equation}
\label{p2}
\phi^2 \sim \xi^3,
\end{equation}
and so on. In general, the number of steps required to cancel $\phi^2$
to an accuracy $\xi^{k}$ is
\begin{equation}
\label{steps}
({\rm number~of~steps}) \sim \xi^{Nk/2}.  
\end{equation}

All the allowed values of $\phi$ near $\phi=0$ have nearly identical
vacuum energies, and the corresponding regions will therefore occupy
equal fractions of the volume in the post-inflationary universe.
The corresponding prior probability for $\phi$ is then simply
proportional to the density of states,
\beq
\label{prob}
{\cal P}_*(\phi)d\phi \propto d\phi/\phi^N.
\eeq

Thus, regions with zero mass and VEV of $\phi$ are maximally probable.
This hierarchy attractor provides a dynamical mechanism for explaining
a zero mass of an interacting scalar without need for supersymmetry.

We note that although the wall charge vanishes as $\phi\to 0$, the
wall tension remains large, $\sigma\sim M_p^3$, and in the limit the
walls become simply domain walls separating degenerate vacua.
Nucleation of such walls is suppressed by a huge
factor \cite{Rama} $\sim \exp (-\pi M_p^2/H^2)$, where $H$ is the
expansion rate during inflation.  This, however, does not change our
conclusions, since eternal inflation provides unlimited time for the
distribution (\ref{prob}) to establish. 

In order to use this "attractor" mechanism for solving the gauge
hierarchy problem, we have to overcome the fact that the attractor
point is at {\it exactly} zero mass and VEV of $\phi$.  In the
following section we will show that curvature corrections to the
potential $V(\phi)$ generally shift the attractor point away from zero
to a small value of $\phi$.

\section{Small Higgs mass from quantum fluctuations}

In the above analysis we have ignored the effects of the gravitational
back reaction on the Higgs mass.  One possible source of this back reaction
is a non-minimal coupling to the curvature,
\begin{equation}
\label{curvature}
|\phi|^2 \, R.
\end{equation}
This will create an additional contribution to the Higgs mass during
inflation,
\begin{equation}
\label{rmass}
\Delta m_{curvature} ^2 \, \sim \, H^2. 
\end{equation}
Even in the absence of such coupling, $\phi$ will get a thermal-type
contribution to its mass due to de Sitter quantum fluctuations.  This
effect is analogous to that of a thermal bath at temperature $T_{GH}
\, \sim \, H$ (Gibbons-Hawking temperature \cite{GH}).  In the domains
where the Higgs VEV drops below $H$, the corresponding contribution to
the Higgs mass is 
\beq 
\Delta m^2_{GH}\sim H^2.  
\label{mGH}
\eeq 
(For $\phi \gg H$, the fields interacting with $\phi$ get masses
greater than $T_{GH}$ and do not contribute to the Higgs potential.)

 To analyze the effect of these contributions on our attractor
mechanism, we shall first consider a simplified picture, where the
expansion rate $H$ remains nearly constant during inflation.  (This
situation is realized in some models of hybrid inflation \cite{hybrid}.)
Due to the mass corrections (\ref{rmass}), (\ref{mGH}),
which we shall assume to be positive, the Higgs VEV
during inflation will not be given by (\ref{vevp}), but will rather be
shifted to
\begin{equation}
\phi^2_{inflationary} \, \sim \, ( m^2 \, -\, F^2/
M_p^2 \, - \Delta m^2),
\end{equation}
where $\Delta m^2 = \Delta m^2_{curvature} \, + \Delta m^2_{GH} \sim
H^2$.  Now, the discussion in the preceeding section indicates that
$\phi$ will be driven not to the point where its post-inflationary
VEV (\ref{vevp}) vanishes, but rather to the point where its {\it
inflationary} VEV vanishes.  That is, most of the space in the
Universe will be occupied by domains where $\phi^2_{inflationary} \,
\sim \, 0$.

 Now, it should be noted that light scalar fields with masses
$m\lsim H$ are subject to large quantum fluctuations during inflation.
Assuming the minimum of the potential is at $\phi=\phi_0$, the
characteristic amplituce of the fluctuations $\delta\phi$ is generally
given by $[V(\phi_0+\delta\phi)-V(\phi_0)]\sim H^4$ \cite{Yokoyama}.
In our case, if the mass of $\phi$ is driven to zero, then $V(\phi)$
is reduced to the quartic term in (\ref{vpf}), and we have
$\delta\phi\sim H$.  This means that the field $\phi$ can be driven to
zero only with an accuracy $\sim {\cal O}(H)$,
\begin{equation}
\label{vevinf}
\phi^2_{inflationary} \, \sim \, ( m^2 \, -\, F^2/
M_p^2 \, - \, {\cal O}(H^2)) \, \sim \, H^2.
\end{equation}
After the end of inflation, the gravitational ($\sim H^2$)
contribution to the mass vanishes, and the VEV is shifted to
\begin{equation}
\label{vevtd}
\phi^2_{today} \, \sim \, ( m^2 \, -\, F^2/ M_p^2 \,)
\, \sim H^2.
\end{equation}

Thus, the post-inflationary Higgs mass will not be exactly zero, but
will rather be comparable to the inflationary Hubble parameter. 
Because of the quantum fluctuations, the infinite peak in the
probability distribution (\ref{prob}) will be smeared, and the
distribution will be nearly flat, ${\cal P}_*(\phi)\approx {\rm
const}$, for $|\phi|\lsim H$.  At larger values of $\phi$, the
probability suppression is at least as strong as in Eq.~(\ref{prob}).
It can be even stronger, due to the effect of differential expansion.
Larger expectation values of $\phi$ correspond to smaller values of
$F$, resulting in a smaller vacuum energy (both because the potential
(\ref{vpf}) is more negative, and because the $F$-field energy density,
$\rho_F\sim F^2$, is smaller).  As a result, the inflationary
expansion rate is lower, which can lead to an exponential suppression
at large $\phi$ \cite{GV2}.

 In order to solve the hierarchy problem, we require that the peak
of the probability distribution is at $\phi\sim 1$~TeV.  Then the
Hubble expansion rate during inflation must be $H\sim 1$~TeV, and the
corresponding vacuum energy density $\sim (10^{11}~{\rm GeV})^4$.
This is a constraint that our mechanism imposes on the inflationary
scenario.

 We now discuss the dynamics of the model in some more detail.
Quantum fluctuations of $\phi$ occur on length and time scales $\delta
l \sim \delta t \sim H^{-1}$.  Fluctuations at the locations of domain
walls cause fluctuations of the wall charge $q$, which in turn cause
variation of the $F$-form in the adjacent domains.  This variation
propagates in the form of waves (see Appendix B), from the walls to
the interior of the domains.  The wavelengths of these waves are
stretched by the exponential expansion of the universe, and as a
result, the form field will vary on an exponentially large scale in
the domain interiors.  Shorter waves, emitted near the end of
inflation, have not travelled far away from the walls.  The sizes of
the domains are huge compared to the present horizon, and it will take
an exponentially long time for the shorter waves to propagate well
into domain interiors.


 After the end of inflation, the Higgs mass takes its
zero-temperature value, and the Higgs rolls away to its new minimum.
At this point, the charge of the walls and the values
of $F$-form in the adjacent regions change, but once again, this change
propagates in the form of waves and is confined to the neighborhood of
the walls.  Moreover, the hierarchy is not destabilized even in
regions affected by the change.
The change of $F$ triggered by the wall is proportional to
the final charge of the wall, which in attractor domains is set by
today's value of the Higgs field $\sim 100$ GeV. Thus, even after the
new value of the field-strength is established, the change with
respect to the inflationary value will be $\Delta F \sim 10^{-17N}\,
M_p^2$. Already for $N=2$, the corresponding change in $\phi$ is
$\Delta \phi\sim 100$~GeV, which does not upset the hierarchy.  For
$N>2$, the back reaction on $\phi$ is negligible.

 Let us finally discuss how the above scenario is modified in more
generic models of inflation, in which the value of $H$ fluctuates
during eternal inflation and then gradually
decreases during a prolonged slow-roll period.  To be specific, we
shall assume that inflation is of the ``new'' type, with expansion
rate $H_{max}\ll M_p$ at the maximum of the potential and $H_{min}$ at
the end of inflation.  Another important parameter is the borderline
value $H_*$ between the regimes of eternal inflation and slow roll.
In the course of eternal inflation, the bubble walls will be exposed
to $H$ in the range $H_*\lsim H \lsim H_{max}$.  Since the
bubble nucleation rate is so low, a typical geodesic will go through
the whole range many times between successive nucleations.  This
suggests that the accuracy with which the attractor mechanism can
drive $\phi_{inflationary}$ to zero cannot exceed $\sim H_{max}$.  The
slow roll period is relatively short and therefore affects only the
immediate vicinity of the bubble walls.  Thus, in order to solve the
hierarchy problem, we have to require $H_{max}\sim 1$~TeV.

\section{Conclusions}

It is usually assumed that the solution to the hierarchy problem
requires introduction of some new physics at low energies. In the
present paper we have provided a counterexample to this statement.  We
have suggested a novel cosmological selection mechanism in which
small scalar masses and VEVs become attractors during the cosmological
evolution.
 
The key idea is that {\it (1)} the scalar mass is dynamically promoted
to a stochastic variable that undergoes discrete jumps during eternal
inflation; {\it (2)} the size of the minimal step is a continuous
function of the "jumping" scalar VEV $\phi$.  That is, the order
parameter in question controls its own steps.  As a result the
probability distribution is sharply peaked around a small value, for
which the step vanishes. We call such a value an "attractor".  The
post-inflationary value of the scalar mass is determined by the
mismatch of masses during and after inflation, due to the
Gibbons-Hawking temperature $T_{GH}\sim H$, and by the magnitude
of quantum fluctuations of $\phi$, also $\sim H$.  Thus, the observed
value of the Higgs mass in our scenario is determined not by
ultraviolet physics, but rather by the maximal inflationary expansion
rate $H_{max}$.  Our scenario requires that $H_{max} \sim 1$~TeV.  It
is obvious that this constraint is not in any respect analogous to the
conventional approaches invoking supersymmetry or some other new
physics around TeV.

Our model involves non-local couplings of the form field $F$, and we
have emphasized that it is not presently clear how this kind of
coupling can be obtained in the low-energy effective theory.  We note
however that this coupling does not seem to have any of the
pathologies usually associated with non-local interactions.  Another
unusual feature of our model is that it has solutions describing waves
of the field $F$ propagating at the speed of light.  This
suggests the presence of massless degrees of freedom, which can
potentially lead to testable predictions.  These issues need further
investigation.

Finally, let us note that the "attractor" mechanism can be combined
with the usual anthropic approach to solving the cosmological constant
problem \cite{GV3}, if one is willing to introduced more three-form
fields.  The main technical problem in this approach has been to
guarantee a sufficiently small step of vacuum energy variation.  With
our attractor mechanism this is trivially achieved, as it generates
branes with tiny charges.  Anthropic selection can also be used
to explain the observed Higgs VEV in models of inflation with
$H_{max}\gg 1$~TeV.


\vspace{0.5cm}

{\bf Acknowledgments}
\vspace{0.1cm} \\

We would like to thank Gregory Gabadadze, Jaume Garriga, 
and Andrei Gruzinov for useful discussions. 
The work of G.D. is supported in
part by a David and Lucile  Packard Foundation Fellowship
for  Science and Engineering,
by Alfred P. Sloan foundation fellowship and by NSF grant
PHY-0070787.  The work of A.V. is supported in part by the NSF.

\appendix

\section{Branes with charges suppressed by symmetries}

In this Appendix we shall give a more detailed discussion of branes
and domain walls with field-dependent three-form charges. 
 Below Planck energies, such branes can be treated
as fundamental objects, or be explicitly constructed as field
theoretic solitonic domain walls, as it was suggested in
\cite{DV}.  The dynamics of
three-form fields in $3+1$ dimensions is in many respects analogous to
the $(1+1)$-dimensional electrodynamics, with electrically charged
particles playing the role of "branes". So we shall first review our
mechanism in a simplified 2-dimensional example, and then generalize
to $(3+1)$ dimensions.  For simplicity, we shall work in Planck units,
and will set all the mass scales equal to one.

The action describing a $(1 + 1)$-dimensional gauge field interacting
with point-like charges can be written as
\begin{equation}
\label{el1}
S_{1+1} \, = \,\int\, d^2x \, F^2 \, + \, q \, \int \, dx^{\mu}\,
A_{\mu},
\end{equation}
where $F$ is the field strength and $x^{\mu}$ is the 
coordinate of a point charge $q$. With $q =$constant, the above
action is gauge invariant under $A_{\mu} \rightarrow A_{\mu} \, + \,
\partial_{\mu}\, \omega$. However, we would like to promote the charge
$q$ to a function of a scalar field $\phi$, that is, $q = \phi^N$.
With this substitution, however, the action (\ref{el1}) is no
longer gauge invariant.  In order to restore the gauge symmetry, we
shall modify it in the following way:
\begin{equation}
\label{gau1}
S_{1+1} \, = \, \int\, d^2x \, F^2 \, + \, \int \,
\phi^N \, dx^{\mu}\, \Pi_{\mu\nu} A^{\nu},
\end{equation}
where $\Pi_{\mu\nu} \, = \, \eta_{\mu\nu} \, - \,
{\partial_{\mu}\partial_{\nu} \over \partial^2}$ is the transverse
projector.  The modification is only significant for a varying
$\phi$. For a constant $q \, = \, \phi^N$, the actions (\ref{gau1}) and
(\ref{el1}) are equivalent.

The generalization of this action to a $(3+1)$-dimensional model with
2-branes is straightforward:
\begin{equation}
\label{gau3}
S_{3+1} \, = \, \int\, d^4x \, F^2 \, + \, \int \,
\phi^N \, d\sigma_{\mu\nu\gamma} \, \Pi^{\mu\mu'}
\Pi^{\nu\nu'}\Pi^{\gamma\gamma'} \, A_{\mu'\nu'\gamma'}
\end{equation}
where $d\sigma_{\mu\nu\gamma}$ is the world-volume element.
This action  is manifestly gauge invariant under (\ref{gauge}). 

 An unusual feature of the actions (\ref{gau1}), (\ref{gau3}) is
that the projection operators $\Pi_{\mu\nu}$ are non-local.  It appears
that such operators cannot be obtained from a local field theory by integrating
out a finite number of heavy particles.  It Ref.~\cite{DV}, the action
(\ref{gau1}) was obtained after integrating over massless fermions in
a $(1+1)$-dimensional model.  However, it is not clear how generic
that model is and whether or not it can be extended to higher
dimensions.  These issues require further study.

As we already mentioned, the branes (or charges) in question can be
regarded as fundamental objects, or as solitons of the effective field
theory.  We shall briefly review the latter possibility.  Following
\cite{DV}, we assume that in the $4D$ low-energy effective theory $F$
is mixed with a certain phase field $a$,
\begin{equation}
\label{af}
 {q \over 2\pi} \, a\, F.
\end{equation}
The interaction (\ref{af}) is invariant under the shift symmetry
\beq
a \rightarrow a\, + \, 2\pi,
\label{shift}
\eeq as well as under gauge transformation of the three-form field
$A_{\mu\nu\alpha}$.  The equation of motion of the three-form field
then demands that the variations of the field strength and $a$ with
respect to any particular coordinate must satisfy
\begin{equation}
\label{daf}
\Delta F\, = \, {q \over 2\pi}\, \Delta a
\end{equation}
Thus, any stable solitonic configuration across which $a$ changes by a
finite amount, inevitably acts as a source for the three-form field.
Such configurations do indeed exist.  Since $a$ is a phase, the
potential of $a$ must respect the shift symmetry (\ref{shift}).  As
long as this is the case, irrespective of the precise form of this
potential, due to topological reasons there must be domain wall
solutions across which $a$ changes by $\Delta a = 2\pi$.  Then,
according to (\ref{daf}), such walls acquire a charge $q$ under $A$.
Thus, the values of the field strength on the two sides of the wall
differ by
\begin{equation}
\label{dfpi}
\Delta F \, = \,  q.
\end{equation}
The charge $q$ defines the minimal step by which $F$ can change from
one region of the Universe to another.

For our purposes, however, $q$ cannot be truly constant, because it
has to depend on  $\phi$.  Also, if we want to think  of $a$ as a 
phase of a certain complex order parameter, 
\beq 
X\, = \, |X|\, {\rm e}^{ia}, 
\eeq 
with a VEV around the Planck scale, $|X| \, \sim  1$, then $q$ will 
depend on $|X|$ as well.  In terms of the fields $X$ and $\phi$, 
the gauge-invariant  coupling (\ref{gau3}) can be
written as
\begin{equation}
\label{PPP}
A_{\mu\nu\gamma}\, \Pi^{\mu\mu'} \, \Pi^{\nu\nu'}\, \Pi^{\gamma\gamma'}\, 
\left ( i \epsilon_{\mu'\nu'\gamma'\alpha}\, (X\partial^\alpha X^\dagger-X^\dagger\partial^\alpha X) \, \phi^N\right )
\end{equation}
This is the lowest possible $a-A$-mixing operator invariant under
gauge (\ref{gauge}) and shift (\ref{shift}) symmetries, and under the
$Z_{2N}$ symmetry which acts on the field $\phi$ as
\begin{equation}
\label{zna}
\phi \, \rightarrow \, {\rm e}^{i{\pi \over N}}\, \phi~.
\end{equation}
The action of this symmetry on the other fields can be defined in 
two alternative ways.  One possibility is to demand
\begin{equation}
a \,  \rightarrow \, -a,~~~ A \, \rightarrow \, \, A,
\end{equation}
or equivalently, $X\rightarrow X^\dagger$.  
Note that the transformation $a\rightarrow -a$ replaces solitons by
anti-solitons and vice versa. 
An alternative choice would be to demand
\begin{equation}
\label{a-}
a \,  \rightarrow \, a,~~~ A \, \rightarrow \, -\, A
\end{equation}
Eq (\ref{PPP}) is invariant under both of these choices.  At low
energies, where $\phi$ and $|X|$ can be treated as constants, the above
coupling reduces to the one of (\ref{af}).  Note that we are not
demanding any approximate or exact $U(1)$ symmetry under the shift $a
\rightarrow a + $constant.  For the existence of the wall, all that we need
is that $|X|\, = \, 0$ be a maximum of the potential.
This suffices to ensure the existence of a stable configuration
across which $a$ changes by $2\pi$.  The corresponding change of $F$
through the wall will be
\begin{equation}
\Delta F \, \sim  \, \phi^N.
\end{equation}


\section{$F$-waves from the walls} 

In $3+1$-dimensions, gauge-invariant free 3-form fields (just as the
$(1+1)$-dimensional electromagnetic field) have no propagating degrees of
freedom, so there are no wave solution.  The situation is different in
our framework, where brane charges depend on the field $\phi$.  At the
end of inflation, $\phi$ rolls away from zero, and the brane charges
change in time.  This change triggers the corresponding change of the
three-form field strength, which propagates away from the brane in the
form of a shock wave.  We shall now discuss this dynamics. 

To illustrate the point, we shall restrict ourselves to a
$(1+1)$-dimensional example. Generalization to $(3+1)$ dimensions is
straightforward.  Labelling the two spacetime coordinates $x^\mu$ by
$t$ and $z$, we shall consider an isolated static
"brane" located at $z=0$.  The gauge-invariant Lagrangian of
interest is
\begin{equation}
\label{lagp}
L\, = \, - {1\over 4} \, F^2 \, + \, A_{\mu}\, \Pi^{\mu 0}\, (\phi^N
\, \delta(z)) \, + |\partial_{\mu}\phi|^2 \, - \, (- \,
m^2_{eff} |\phi|^2 \, + \, |\phi|^4),
\end{equation}
where $m_{eff}$ is the effective mass of $\phi$ after inflation, and
we have ignored higher-order terms in the potential of $\phi$.
When $\phi$ rolls away from $\phi = 0$, the change of the brane charge
induces a back reaction on $\phi$. This back reaction is
suppressed by the brane charge $\sim \phi^N$.  Since in the domains of
interest, the final VEV of $\phi$ is very small, the back reaction
on $\phi$ is negligible, and we shall ignore it.

Thus, we shall study the dynamics of a gauge field in the background
of a time-dependent charge. This dynamics is governed by the following
equation
\begin{equation}
\label{feq}
\partial^{\mu}\, F_{\mu\nu} \, = \, \Pi_{\nu 0}\, (\phi^N(t)\, \delta
 (z)) \, = \, \eta_{\nu 0} \, \phi^N(t) \, \delta(z)\, -
 \int \, {dp^2 \over (2\pi)^2} {p_{\nu}p_0 \over p^2} \, {\rm
 e}^{-ipx} \, \tilde{\phi_{N}}(p_0),
\end{equation}
where $p$ is the two-momentum, and $\tilde{\phi_{N}}(p_0)$ is the
Fourier-transform of $\phi^N(t)$.  A straightforward
integration gives the following two equations
\begin{eqnarray}
 \partial_z \, F_{10} \, & = & \, {1
\over 2} \, \partial_{z} \left ( \, \phi^N(t\, -\, z) \, \theta
(z) \, - \, \phi^N(t\, + \, z) \, \theta (-z) \, \right),\\ 
\partial_t \, F_{10} \, & = & \, \, {1 \over 2} \, \partial_{t}
\left ( \, \phi^N\, (t\, -\, z) \, \theta (z) \, - \, \phi^N\,
(t\, + \, z) \, \theta (-z) \, \right ),
\end{eqnarray}
which are solved by
\begin{equation}
\label{wave}
 F_{10} \, = \, \, {1 \over 2} \, \left (\, \phi^N\, (t\, -\, z) \,
 \theta (z) \, - \, \phi^N\, (t\, + \, z) \, \theta (- z)
 \,\right ).
\end{equation}
This solution describes waves propagating away from the brane in two
opposite directions.  From this solution it is also obvious that after
the transition is finished and $\phi$ assumes a constant value
$\phi_{today}$, the change of $F$ across the brane is given by
\begin{equation}
\label{ffinal}
\Delta F \, = \, \phi^N_{today}.
\end{equation}

\end{document}